\documentclass[journal]{IEEEtran}
\usepackage{amssymb,stmaryrd,amsmath,amsfonts,rotating}%amsfonts,amsthm
\usepackage[vflt]{floatflt} \usepackage{epic}
\usepackage{color}
\usepackage{bm}
\newcommand{\ttz}{\text{\tt 0}}
\newcommand{\tto}{\text{\tt 1}}
\newcommand{\tte}{\text{\tt ?}}
\newcommand{\NC}{V_{\mathrm{c}}}
\newcommand{\Vt}{V_{\mathrm{t}}}
\newcommand{\Vp}{V_{\mathrm{p}}}
\newcommand{\hL}{\hat{L}}

\newcommand{\GL}{\mathrm{GL}}
\newcommand{\Da}{D_{2}}
\newcommand{\Va}{V_{2}}
\newcommand{\Dch}{D_{\mathrm{0}}}
\newcommand{\Vch}{V_{\mathrm{0}}}
\newcommand{\Dex}{D_{\mathrm{1}}}
\newcommand{\Vex}{V_{\mathrm{1}}}
\newcommand{\dl}{d_{\mathrm{l}}}
\newcommand{\dr}{d_{\mathrm{r}}}
\newcommand{\dg}{d_{\mathrm{g}}}
\newcommand{\dgbc}{\genfrac[]{0pt}0} %% Gaussian binominal coefficient for display style
\allowdisplaybreaks
\begin{document} 
\title{Spatially-Coupled Binary MacKay-Neal Codes for Channels with Non-Binary Inputs and Affine Subspace Outputs}
\author{
Kenta Kasai, Takayuki Nozaki and Kohichi Sakaniwa\\
\authorblockA{
Dept. of Communications and Integrated Systems, 
Tokyo Institute of Technology, 152-8550 Tokyo, Japan.\\
Email: {\tt \{kenta,nozaki,sakaniwa\}@comm.ss.titech.ac.jp}} 
}

\maketitle
\begin{abstract}
We study LDPC codes for  the channel with $2^m$-ary input $\underline{x}\in \mathbb{F}_2^m$ and 
output $\underline{y}=\underline{x}+\underline{z}\in \mathbb{F}_2^m$. 
The receiver knows a subspace $V\subset \mathbb{F}_2^m$ from which $\underline{z}=\underline{y}-\underline{x}$ is uniformly chosen.
Or equivalently, the receiver receives an affine subspace $\underline{y}-V$ where $\underline{x}$ lies. 
We consider a joint iterative decoder involving the channel detector and the LDPC decoder. 
The decoding system considered in this paper can be viewed as a simplified model of 
the joint iterative decoder over non-binary modulated signal inputs e.g., $2^m$-QAM. 
We evaluate the performance of binary spatially-coupled MacKay-Neal codes by density evolution. 
The iterative decoding threshold is seriously degraded by increasing $m$. 
EXIT-like function curve calculations reveal that this degradation is caused by wiggles and can be mitigated by 
increasing the randomized window size. 
The resultant iterative decoding threshold values are very close to the Shannon limit. 
\end{abstract}
%%%%%%%%%%%%%%%%%%%%%%%%%%%%%%%%%%%%%%%%%%
\section{Introduction}
%%%%%%%%%%%%%%%%%%%%%%%%%%%%%%%%%%%%%%%%%%
Spatially-coupled (SC) low-density parity-check (LDPC) codes attract much attention due to their capacity-achieving performance and a memory-efficient sliding-window decoding algorithm.
The studies on SC-LDPC codes date back to the invention of convolutional LDPC codes by Felstr{\"o}m and Zigangirov \cite{zigangirov99}. 
They introduced a construction method of $(\dl, \dr)$-regular convolutional LDPC codes from $(\dl, \dr)$-regular block LDPC codes \cite{mct}. 
The convolutional LDPC codes exhibited better decoding performance than the underlying block LDPC codes under a fair comparison with respect to the code length. 
Lentmaier {\it et al.}~ observed that (4,8)-regular convolutional LDPC codes exhibited the decoding performance surpassing the belief propagation (BP) threshold of (4,8)-regular block LDPC codes \cite{lentmaier_II}. 
Further, the BP threshold coincides with the maximum a posterior (MAP) threshold of the underlying block LDPC codes with a lot of accuracy. 
Constructing convolutional LDPC codes from a block LDPC code improves the BP threshold up to the MAP threshold of the underlying codes. 

Kudekar {\it et al.}~ named this phenomenon  ``threshold saturation'' and proved rigorously for the binary-input erasure channel (BEC) \cite{5695130} and the binary-input memoryless output-symmetric (BMS) channels. \cite{2012arXiv1201.2999K}. 
In the limit of large $\dl,\dr,L$ and $w$, the SC-LDPC code ensemble $(\dl,\dr,L,w)$ \cite{5695130} was shown to {\em universally} achieve the Shannon limit of BMS channels under BP decoding. 
This means the transmitter does not need detail statistics of the channel but needs to know only the channel capacity. 
Such universality is not supported by other efficiently-decodable capacity-achieving codes, e.g., polar codes \cite{5075875} and irregular LDPC codes \cite{richardson01design}.
According to the channel, polar codes need  frozen bit selection \cite{5075875} and irregular LDPC codes need optimization of degree distributions. 
We note that recently Aref and Urbanke proposed SC rateless codes \cite{ITW_AREF_URBANKE} which  are conjectured to universally achieve the capacity of BMS channels without knowing even the capacity of the channel at the transmitter. 

MacKay-Neal (MN) codes \cite{mn_code} are non-systematic two-edge type LDPC codes \cite{met,mct}. 
The MN  codes are conjectured to achieve the capacity of BMS channels under maximum likelihood decoding. 
Murayama {\it et al.}~\cite{PhysRevE.62.1577} and Tanaka and Saad \cite{TanakaSaad2003} reported the empirical evidence of the conjecture for BSC (Binary Symmetric Channel) and AWGN (Additive White Gaussian Noise) channels, respectively 
by a non-rigorous but powerful statistical mechanics approach known as {\em replica method}. 
In \cite{HSU_MN_IEICE}, Kasai and Sakaniwa presented a spatial coupling method of SC-MN codes. 
Empirical results showed that SC-MN codes with bounded density achieve the capacity of the BEC. 
It was observed that the SC-MN codes have the BP threshold close to the Shannon limit.

In this paper, we study coding over the channel with $2^m$-ary input $\underline{x}\in \mathbb{F}_2^m$ and output $\underline{y}\in \mathbb{F}_2^m$. 
The receiver knows a subspace $V\subset \mathbb{F}_2^m$ from which $\underline{z}=\underline{y}-\underline{x}$ is uniformly chosen.
Or equivalently, the receiver receives an affine subspace $\underline{y}-V:=\{\underline{y}-\underline{z}\mid\underline{z}\in V\}$ in which the input $\underline{x}$ is compatible. 
This channel model is used in the decoding process for network coding \cite{DBLP:journals/corr/abs-0711-3935} after estimating noise packet spaces. 
In \cite{DBLP:journals/corr/abs-0711-3935}, non-binary LDPC codes are used, which results in high-decoding complexity $O(m^3)$ at each  channel factor node and parity-check nodes. 
We do not use non-binary codes but binary codes for coding the non-binary input channel. 
Furthermore, we consider the joint iterative decoding between the channel detector and the code decoder. 
The channel detector calculates log likelihood ratio (LLR) of the transmitted bits from a channel output and messages from the BP decoder. 
Such a decoding system can be viewed as the simplest model of joint iterative decoders which involve the channel detector of non-binary modulation (e.g., $2^m$-QAM (quadrature amplitude modulation)) and the BP decoder for LDPC codes.

The aim of this paper is to evaluate the joint iterative decoding performance of binary SC-MN codes over the non-binary input channels. 
To this end, we use density evolution (DE) analysis for joint iterative decoding of binary SC-MN codes. 
EXIT-like function curve calculations reveal that BP threshold values are very close to the Shannon limit. 
%%%%%%%%%%%%%%%%%%%%%%%%%%%%%%%
\section{Channel Model}
%%%%%%%%%%%%%%%%%%%%%%%%%%%%%%%
In this paper, we consider channels with input $\underline{x}\in \mathbb{F}_2^m$ and output $\underline{y}=\underline{x}+\underline{z}\in \mathbb{F}_2^m$, where  $\underline{z}\in \mathbb{F}_2^m$ is uniformly distributed in a linear subspace $V\subset\mathbb{F}_2^m$. We refer to this subspace as noise subspace. 
The dimension of $V$ is distributed as $\Pr(\dim(V)=d)=:p_d$. 
Given the dimension $d$, $V$ is also uniformly distributed, i.e., $\Pr(V=v|\dim(V)=d)=1/\dgbc{m}{d}$, 
where $\dgbc{m}{d}=\prod_{l=0}^{d-1}\frac{2^m-2^l}{2^d-2^l}$
is a 2-Gaussian binomial. 
The normalized capacity per input bit is given by 
\begin{align*}
 {1\over m}\max_{p(\underline{X})}(I(\underline{X};\underline{Y})) ={1\over m}\max_{p(\underline{X})}(H(\underline{Y})-H(\underline{Y}|\underline{X}))
\end{align*}
The latter part equals to $H(\underline{Z})$ which is independent of $p(\underline{X})$ and the former part is maximized when $\underline{X}$ is uniformly distributed. 
Hence, it follows that the normalized channel capacity is given as $1-\sum_{d=0}^m{d\over m}p_d.$

We consider two types of smooth channels defined by $p_d$. 
The simplest dimension distribution is deterministic one. 
The channel  W($m,w$)  for $w\in \{0,1,\dotsc,m\}$ is defined by 
 \begin{align*}
 &p_d:=\left\{
\begin{array}{ll}
1&(d=w), \\
0&(d\neq w).
\end{array}
\right.
 \end{align*}
%---------------------------
W($m,w$) for large $m$ was used in a decoding process of the network coding scenario \cite{DBLP:journals/corr/abs-0711-3935}. 
In \cite{DBLP:journals/corr/abs-0711-3935}, the data part of each packet is represented as $\underline{x}\in\mathbb{F}_2^m$. 
Packets are coded by non-binary LDPC codes whose parity-check coefficients are in the general linear group $\GL(m,\mathbb{F}_2)$. 
The noise subspace $V$ is estimated by padding zero packets and using Gaussian elimination. 
Note that, in this paper, unlike \cite{DBLP:journals/corr/abs-0711-3935} we use binary LDPC codes for coding the $2^m$-ary input channel. 
Note also that, our DE calculation is currently limited for small $m$. 

In order to evaluate the performance of a given coding and decoding system, we prefer smoothly parametrized channels. 
Instead of W($m,w$), let us consider CD($m,\epsilon$) as a smoothly parametrized version of W$(m,w)$, where CD stands for concentrated dimension. 
 \begin{align*}
 &p_d:=\left\{
\begin{array}{rl}
1-\epsilon m+ \lfloor\epsilon m\rfloor&(d=\lfloor\epsilon m\rfloor), \\
\epsilon m- \lfloor\epsilon m\rfloor&(d=\lfloor\epsilon m\rfloor+1),\\
0&(\text{otherwise}).\\
\end{array}
\right.
 \end{align*}
For $m\epsilon=w$, W($m,w$) is identical to CD($m,\epsilon$). 
The normalized capacity of CD($m,\epsilon$) is given by $1-\epsilon$. 

Another interesting non-binary input channel is BD($m,\epsilon$) whose dimension distribution is given by the binomial distribution
$p_d:=\binom{m}{d}\epsilon^{d}(1-\epsilon)^{m-d},$
where BD stands for binomial distribution. 
This channel is realized by multiplying random $m\times m$ binary non-singular matrix $A$ by the input bits $\underline{x}$ before transmitting through the binary erasure channel with erasure probability $\epsilon$ \cite{4111/THESES}. 
The affine subspace is given by $\{\underline{x}:A_E\underline{x}_E^T=A_{E^\mathrm{c}}\underline{x}^T_{E^\mathrm{c}}\}$, where $E\subset \{1,\dotsc,m\}$ 
is the set of erased output indices, $A_E$ denotes the submatrix of $A$ indexed by the elements of $E$, and $\underline{x}_{E}$ is the corresponding subvector. 
The normalized capacity of BD($m,\epsilon$) is given also as $1-\epsilon$. 

%%%%%%%%%%%%%%%%%%%%%%%%%%%%%%%%%%
% FIGURE
%%%%%%%%%%%%%%%%%%%%%%%%%%%%%%%%%%
\begin{figure*}[t]
\setlength{\unitlength}{1.0bp}%
\begin{picture}(200,150)(0,0)
\put(0,0){
\put(0,10){\includegraphics[width=0.7\textwidth]{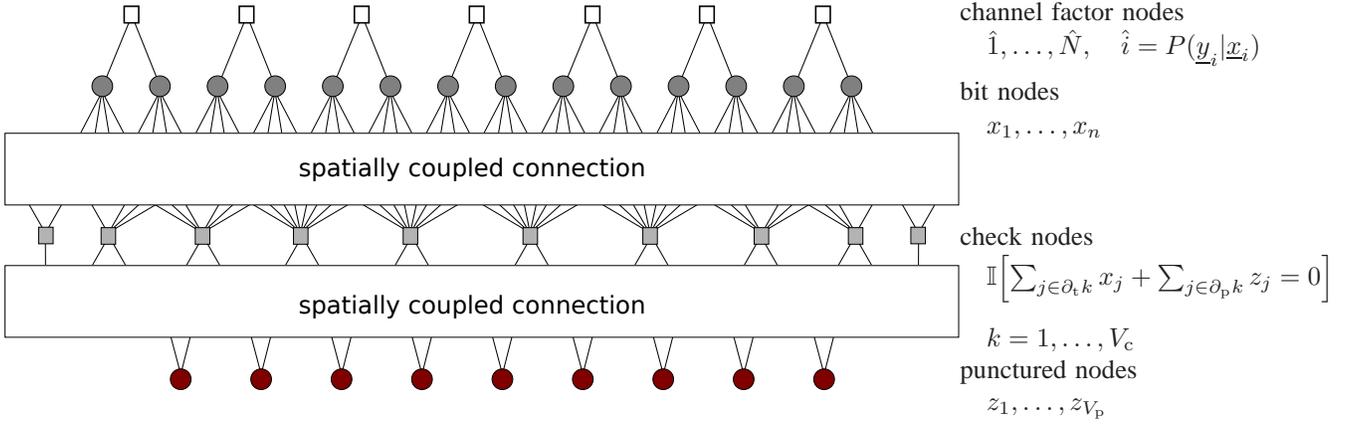} }
\put(360,150){{channel factor nodes }}
\put(370,137){$\hat{1},\dotsc,\hat{N}, \quad \hat{i}=P(\underline{y}_i|\underline{x}_i)$}
\put(360,120){bit nodes}
\put(370,107){$x_1,\dotsc,x_n$}
\put(360,65){check nodes}
\put(370,50){$\mathbb{I}\Bigl[\sum_{j\in\partial_{\mathrm{t}} k}x_j+\sum_{j\in\partial_{\mathrm{p}} k}z_j=0\Bigr]$}
\put(370,27){$k=1,\dotsc,\NC$}
\put(360,15){punctured nodes}
\put(370,3){$z_1,\dotsc,z_{\Vp}$}
}
\end{picture}
\caption{Factor graph representation of  SC-MN codes with $\dl=4, \dr=2, \dg=2$ over CD($m=2,\epsilon$) or BD($m=2,\epsilon$).}
\label{133609_4Feb12}
\end{figure*}
%%%%%%%%%%%%%%%%%%%%%%%%%%%%%%%
\section{Spatially-Coupled MacKay-Neal Codes}
%%%%%%%%%%%%%%%%%%%%%%%%%%%%%%%
Let $N$ denote the number of channel uses for one codeword, in other words, at each channel use, an $m$-bit symbol is transmitted through the channel. 
The total number of transmitted bits are $n:=mN$. We use SC-MN codes of length $n$. 

In this section, we define binary SC-MN codes.  
% and non-binary SC-MN codes. 
% Each symbol $\underline{x}_i$ of the non-binary SC-MN codes is in  $\underline{x}_i\in\mathbb{F}_2^m$ 
% and each parity-check equation 
% \begin{align*}
%  &\sum_{j}h_{i,j}\underline{x}_{j}=\underline{0} \in \mathbb{F}_2^m
% \end{align*}
% has coefficients $h_{i,j}$ over $\GL(\mathbb{F}_2,m)$. 
%-------------------------------------------------
%\section{Binary Spatially-Coupled MacKay-Neal Codes}
%-------------------------------------------------
We define a $(\dl, \dr, \dg, L, w)$ SC-MN code ensemble as follows. 
The Tanner graph of a code in the $(\dl, \dr, \dg, L, w)$ SC-MN code ensemble is constructed as follows. 
At each section $i\in\mathbb{Z}:=\{0,\pm 1,\pm 2,\dotsc\}$, consider
$\frac\dr\dl M$ bit nodes of degree $\dl$, $M$ bit nodes of degree $\dg$. We refer to those two types of bit nodes as type 1 and type 2, respectively. 
Additionally, at each section $i$,  consider $M$ check nodes which adjacent to $\dr$  bit nodes of type 1 and $\dg$  bit nodes of type 2. 
Connect randomly these nodes in such a way that for $i\in \mathbb{Z}$ and $j=0,\dotsc, w-1$, 
bit nodes of type 1 at section $i$ and check nodes at section $i+j$ are connected with $\frac{\dr M}{w}$ edges and 
bit nodes of type 2 at section $i$ and check nodes at section $i+j$ are connected with $\frac{\dg M}{w}$ edges. 
Shorten the bit nodes of type 1 and 2 at section $|i|>L$, in other words, set the bits to zero and do not transmit them. 
Puncture the bit nodes of type 1 at section $|i|\le L$, in other words, the bits are not transmitted. 
Hereafter, we call bit nodes of type 1 and 2, punctured and transmitted bits, respectively. 
Note that this code ensemble is nicely represented as LDPC codes defined by a joint degree distribution \cite{Detailed_IEICE}.
The definition of the $(\dl, \dr, \dg, L, w)$ SC-MN code ensemble is based on that of $(\dl,\dr,L,w)$ randomized SC-LDPC code ensemble. 
For more details on $(\dl,\dr,L,w)$ randomized SC-LDPC code ensemble, we refer the readers to \cite[Section II.B]{5695130}.

Denote the number of transmitted bit nodes, punctured nodes by $\Vt$, $\Vp$, respectively.
$ n=\Vt=\hL M, \quad \Vp=\frac{\dr}{\dl}\hL M,$ where $\hL:=2L+1$. 
The number of check nodes of degree at least 1, denoted by $\NC$, can be counted by the same way as 
in \cite[Lemma 3]{5695130} as follows. 
$ \NC=M[2L-w+2\sum_{i=0}^w(1-(\frac{i}{w})^{\dr}(\frac{i}{w})^{\dg})].$
The design coding rate $R^{\mathrm{MN}}(\dl,\dr,L,w)$ is given by 
$ R^{\mathrm{MN}}(\dl,\dr,L,w):=\frac{\Vt+\Vp-\NC}{V_\mathrm{t}}
 =\frac{\dr}{\dl}+\frac{1+w-2\sum_{i=0}^w(1-(\frac{i}{w})^{\dr}(\frac{i}{w})^{\dg})}{\hL}
=\frac{\dr}{\dl} \quad (\hL\to\infty).
$
%-------------------------------------------------
%\section{Non-Binary Spatially-Coupled MacKay-Neal Codes}
%-------------------------------------------------
%%%%%%%%%%%%%%%%%%%%%%%%%%%%%%%
\section{Decoding Algorithm}
%%%%%%%%%%%%%%%%%%%%%%%%%%%%%%%
In this section, we define a factor graph \cite{910572} for decoding the SC-MN code over the channel. 
Denote a $(\dl, \dr, \dg, L, w)$ SC-MN code by $C$. 
Denote the transmitted code symbols, punctured bits and the received symbols by 
$\bm{x}=(x_1,\dotsc,x_{mN})=:(\underline{x}_1,\dotsc,\underline{x}_N)$, $\bm{z}=(z_1,\dotsc,z_{\Vp})$ and 
$\bm{y}=(\underline{y}_1,\dotsc,\underline{y}_N)$, respectively.

We adopt the sum-product algorithm \cite{910572} over a factor graph defined by the following factorization of $P(\bm{x},\bm{z}|\bm{y})$.
\begin{align}
& P(\bm{x},\bm{z}|\bm{y})=P(\bm{y}|\bm{x},\bm{z})P(\bm{x},\bm{z})/P(\bm{y}),\nonumber\\
&P(\bm{y}|\bm{x},\bm{z})=\prod_{i=1}^NP(\underline{y}_i|\underline{x}_i),\nonumber\\
&P(\underline{y}_i|\underline{x}_i)=\left\{\begin{array}{cl}
2^{-\dim(V_i)}& (\underline{y}_i-\underline{x}_i\in V_i),\label{112503_30Jan12}\\
0& (\text{otherwise}),
\end{array}\right.\\
& P(\bm{x},\bm{z})=\prod_{k=1}^{\NC} \mathbb{I}[\sum_{j\in\partial_{\mathrm{t}} k}x_j+\sum_{j\in\partial_{\mathrm{p}} k}z_j=0]/\#C,\label{112614_30Jan12}
\end{align}
where $\partial_{\mathrm{t}} k$ (resp.  $\partial_{\mathrm{p}} k$) is the set of indices for transmitted (resp. punctured) bit nodes adjacent to the $k$-th check node of $C$, and $\#C$ represents the size of $C$. 
Figure \ref{133609_4Feb12} shows an example of a factor graph 
representation of  SC-MN codes with $\dl=4, \dr=2, \dg=2$ over CD($m=2,\epsilon$) or BD($m=2,\epsilon$).

The calculation for message $\mu_{i\to im+j}(x_{im+j})$ from the factor node $\hat{i}$ of \eqref{112503_30Jan12} to the variable node $x_{im+1}$ is given by 
\begin{align*}
&\mu_{\hat{i}\to im+1}(x_{im+1}) \propto\\
&\sum_{x_{im+2},\dotsc, x_{(i+1)m}\in\mathbb{F}_2}\hspace{-0.5cm}P(\underline{y}_i|\underline{x}_i)\prod_{j=2}^{m}\mu_{im+j\to \hat{i}}(x_{im+j}), 
\end{align*}
where $\mu_{im+j\to \hat{i}}(x_{im+j})$ is the sum-product message from the variable node of $x_{im+1}$ to $\hat{i}$. 
At the beginning of the algorithm, $(\mu_{im+j\to \hat{i}}(0),\mu_{im+j\to \hat{i}}(1))$ is set to $(1/2,1/2)$. 
From \cite[Chap. 2]{mct}, we know that if $(\mu_{im+j\to \hat{i}}(0),\mu_{im+j\to \hat{i}}(1))$ is either $\ttz:=(1,0)$, $\tto:=(0,1)$ or $\tte:=(1/2, 1/2)$, $\mu_{\hat{i}\to im+j}(x_{im+j})$ is also either \ttz,\tto, or \tte. 
This message calculation at $\hat{i}$ is efficiently accomplished by Gaussian elimination with $O(m^3)$ calculations. 
Since the factor node for each factor in \eqref{112614_30Jan12} is a check node. 
Hence, the messages stay in \{\ttz, \tto, \tte\}. 

%%%%%%%%%%%%%%%%%%%%%%%%%%%%%%%%%%%%%%%%%%%%%%%%%%%%
\section{Density Evolution Analysis}
%%%%%%%%%%%%%%%%%%%%%%%%%%%%%%%%%%%%%%%%%%%%%%%%%%%%
In this section, we derive DE equations for the sum-product algorithm over the factor graph defined in the previous section.%-------------------------------------------
\subsection{Factor Node of Channel}
%-------------------------------------------
The messages stay in \{\ttz, \tto, \tte\} during the entire decoding process. DE update equation for the check node is obvious. 
First, we will  focus on deriving the DE for the factor node $\hat{1}$ of \eqref{112503_30Jan12}. 

Rathi \cite{4111/THESES} developed the DE of non-binary LDPC codes for the BEC. 
The DE allows us to  track probability mass functions of the dimension of the linear subspaces. 
For $\ell\ge 0$, the DE tracks the probability vectors $\underline{P}^{(\ell)}=(P_0^{(\ell)},\dotsc, P_m^{(\ell)})\in [0,1]^{m+1}$
 which are referred to as {\itshape densities}. 

Without loss of generality, we can assume all-zero codewords were sent.
We fix the first one of $m$ transmitted bit nodes $1,\dotsc,m $ adjacent to the factor node $\hat{1}$ and derive the probability that the outgoing message $\mu$ along the edge is erasure. 
There are $m-1$ incoming messages  to $\hat{1}$.
Denote the probability that a randomly picked message from the transmitted bit nodes are erased by $z$. 
Denote the indices of known messages by $\mathcal{K}$. Note that $1\notin \mathcal{K}$. 
Define a subspace $V_1=\{\underline{x}=(x_1, \dotsc, x_m)\in\mathbb{F}_2^m\mid x_i=0, i\notin \mathcal{K}\}$ and denote $\Dex:=\dim(V_1)$. We have $\#\mathcal{K}=m-\Dex.$
Then, it follows that \begin{align*}
 &P_{\Dex}^{(z)}(i+1)=
 \left\{\begin{array}{ll}
 \binom{m-1}{i}z^i(1-z)^{m-i-1}& (0\le i\le m-1)\\
 0 &(i=-1).
 \end{array}\right.
\end{align*}
Let $\Vch\subset\mathbb{F}_2^m$ be the noise subspace which is known at $\hat{1}$. 
The outgoing message is not erased if and only if all vectors in $\Va:=\Vex\cap \Vch$ have 0 at the first index, in precise, 
$\forall (x_1,\dotsc,x_m)\in \Va, x_1=0$. 
From \cite{4111/THESES}, 
the probability $P_{\Da|\Dch,\Dex}(k|j, i)$ that $\Va$ has dimension $k$ given  $\Vex$ has dimension $i$ and $\Vch$ has dimension $j$ is given as follows. 
\begin{align*}
 &P_{\Da|\Dch,\Dex}(k|j, i)=2^{(i-k)(j-k)}{\dgbc{i}{k}\dgbc {m-i} {j-k}  }{\dgbc m j }^{-1}.
\end{align*}

Since bits whose indices in $\mathcal{K}$ are known, i.e., \ttz, $\Da$ is distributed over index support $\{1,\dotsc,m\}\setminus\mathcal{K}$.
The outgoing message $\mu$ is known if the support of $\Va$ does not include $1$. 
By counting the number for subspace of dimension $i$ and $i-1$ over the support of size $k$, 
 we obtain the probability $P_{\mu|\Da,\Dex}(\tte|k,i)$ that outgoing message is \tte\ given that 
$\Va$ has dimension $k$ and $\Vex$ has dimension $i$ is given as follows. 
\begin{align*}
 &P_{\mu|\Da,\Dex}(\tte|k,i)=1-{\dgbc{i-1}{k}}{\dgbc{i}{k}}^{-1}.
\end{align*}
Therefore, we have the probability $f_{\epsilon,m}(z):=P(\mu=\tte)$ that outgoing the message is \tte\ as follows. 
\begin{align*}
&f_{\epsilon,m}(z)=\sum_{k=0}^{m}\sum_{i=k}^{m}P_{\mu|\Da,\Dex}(\tte|k,i)P_{\Da,\Dex}(k,i),\\
&P_{\Da,\Dex}(k,i)=\sum_{j=k}^{m-i+k}P_{\Da|\Dch,\Dex}(k|j,i)P_{\Dch}(j)P_{\Dex}^{(z)}(i).
\end{align*}
%---------------------------------------
\subsection{Density Evolution for Spatially-Coupled MacKay-Neal Codes}
\label{192913_3Feb12}
%---------------------------------------
Let $p_i^{(\ell)}$ and $q_i^{(\ell)}$ be the erasure probability of messages emitting from punctured bit and transmitted bit nodes to check nodes, respectively, at section $i$ at the $\ell$-th round of BP decoding in the limit of large $M$.
Similarly, let $z_i^{(\ell)}$ be the erasure probability of messages emitting from channel factor nodes to punctured bit nodes at section $i$ at the $\ell$-th round of BP decoding in the limit of large $M$.
DE update equations of the randomized $(\dl, \dr, \dg, L, w)$ SC-MN code are given as follows. 
For $|i|>L$ and $\ell\ge 0$, $p_i^{(\ell)}=q_i^{(\ell)}=0$. 
For $|i|\le L$, $p_i^{(0)}=q_i^{(0)}=1$. 
For $|i|\le L$, 
\begin{align*}
&p_i^{(\ell+1)}=\\
&\bigl(\frac{1}{w}\sum_{j=0}^{w-1}[1-(1-\sum_{k=0}^{w-1}\frac{p_{{i+j}-k}^{(\ell)}}{w})^{\dr-1}(1-\sum_{k=0}^{w-1}\frac{q_{{i+j}-k}^{(\ell)}}{w})^{\dg}]\bigr)^{\dl-1},\\
&s_i^{(\ell)}=\frac{1}{w}\sum_{j=0}^{w-1}[1-(1-\sum_{k=0}^{w-1}\frac{p_{{i+j}-k}^{(\ell)}}{w})^{\dr}(1-\sum_{k=0}^{w-1}\frac{q_{{i+j}-k}^{(\ell)}}{w})^{\dg-1}],\\
&q_i^{(\ell+1)}=f_{\epsilon,m}(z_i^{(\ell)})\cdot(s_i^{(\ell)})^{\dg-1}, \quad z_i^{(\ell)}=(s_i^{(\ell)})^{\dg}.
\end{align*}

%------------------------------------------
\subsection{Threshold Values}
%------------------------------------------
We define the BP threshold value  as $ \epsilon^{\ast}:=\sup\{\epsilon\in[0,1]\mid\lim_{\ell\to\infty}p_i^{(\ell)}=0, i=-L,\dotsc,+L\},$
for the channel CD($m,\epsilon$) or BD($m,\epsilon$).
In words, the SC-MN codes enable reliable transmissions over the channel CD($m,\epsilon$) or BD($m,\epsilon$) if $\epsilon<\epsilon^{\ast}$. 

Threshold values $\epsilon^{*}(\dl,\dr,L,w)$ for $(\dl=4, \dr=2, \dg=2, L, w)$ SC-MN codes are numerically calculated and listed in Tab.~\ref{142501_1Feb12}. 
As was also observed in the case of BEC \cite{HSU_MN_IEICE,5695130}, we also observed the threshold values remain almost the same for large $L$ e.g. 200 for fixed $\dl,\dr,\dg,w$. 
Since the rate of $(\dl=4, \dr=2, \dg=2, L, w)$ SC-MN codes converges to $\dr/\dl=1/2$ in the limit of large $L$, we ignore the rate-loss.
As can be seen, the threshold values are very close to the Shannon limit $\epsilon=1/2$ of  $(\dl=4, \dr=2, \dg=2, L=\infty, w<\infty)$ SC-MN codes.
Interestingly, threshold values are degraded as $m$ increases. 
%This degradation is caused by wiggles as seen also in the case of BEC \cite{5695130}. 
The gap to the Shannon limit seems to grow exponentially with $m$. 
%%%%%%%%%%%%%%%%%%%%%%%%%%%%%%%%%
%FIRUGRE comparison L=10, L=20
%%%%%%%%%%%%%%%%%%%%%%%%%%%%%%%%%
\begin{figure}
\setlength{\unitlength}{1.0bp}%
\begin{center}
 \begin{picture}(180,130)(20,20)
 \put(0,0)
 {
 \put(0,10){\includegraphics[width=0.43\textwidth]{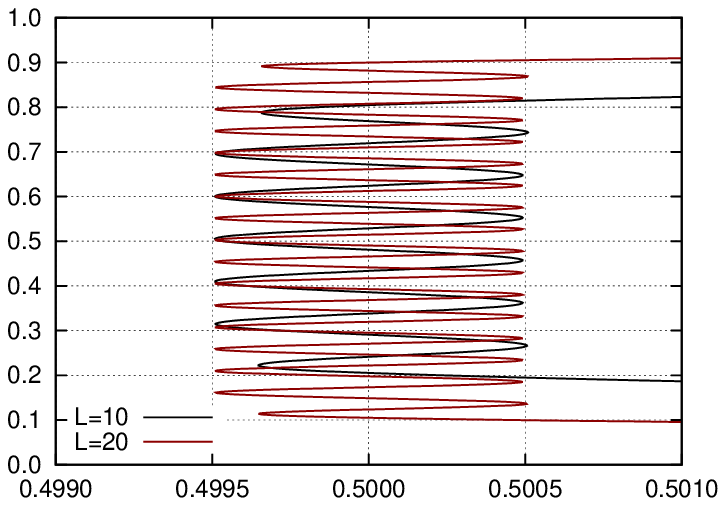} }
 \put(-5, 90){\rotatebox{90}{$h^{\mathrm{EBP}}(\epsilon)$}}
 \put(130,10){$\epsilon$}
 }
 \end{picture}
\end{center}
\caption{
EBP EXIT-like curve of $(\dl=4, \dr=2, \dg=2, L, w=2)$ SC-MN codes with $L=10$ and $L=20$. 
The channel is CD($m=2,\epsilon$).
The threshold for each code is given by erasure probability $\epsilon$ (X-axis) of the leftmost point on the curve. 
The threshold values are almost the same around $\epsilon=0.499509$ for $L=10$ and $L=20.$
}
\label{195027_3Feb12}
\end{figure}
%%%%%%%%%%%%%%%%%%%%%%%%%%%%%%%%%%%%%%%%
% Threshold table for CD
%%%%%%%%%%%%%%%%%%%%%%%%%%%%%%%%%%%%%%%%
\begin{table}[t]
\begin{center}
 \caption{Threshold values $\epsilon^{*}(\dl,\dr,L,w)$ for $(\dl=4, \dr=2, \dg=2, L, w)$ SC-MN codes for CD($m,\epsilon$) and BD($m,\epsilon$). }
 \label{142501_1Feb12}
 \begin{tabular}{c|c|c||c|c}
 &\multicolumn{2}{c||}{CD($m,\epsilon$)}&\multicolumn{2}{c}{BD($m,\epsilon$)}\\\hline
 $m$&$L=10$&$L=20$&$L=10$&$L=20$\\
    &$w=2$&$w=3$&$w=2$&$w=3$\\\hline
 1& 0.49998527&0.49999998& 0.49998527&0.49999998\\
 2& 0.49950900&0.49999936& 0.49987196&0.49999983\\
 3& 0.49913150&0.49999850& 0.49954538&0.49999942\\
 4& 0.49714179&0.49998625& 0.49885380&0.49999741\\
 5& 0.49566948&0.49997243& 0.49768392&0.49999097\\
 6& 0.49166023&0.49989196& 0.49596851&0.49997517
 \end{tabular}
\end{center}
\end{table}

%%%%%%%%%%%%%%%%%%%%%%%%%%%%%%%%%%%
\section{EXIT-like Function}
%%%%%%%%%%%%%%%%%%%%%%%%%%%%%%%%%%%
In this section we investigate the reason of threshold degradation by an EXIT-like function. 
By ``EXIT-like'', we mean that the area theorem \cite{mct} does not necessarily hold. 
However, EXIT-like function allows us to understand how the decoding performance is affected by the fixed points of DE. 

Consider fixed points of the DE system described in Section \ref{192913_3Feb12}, i.e., 
$(\underline{p}:=(p_{i})_{i=-\infty}^{\infty},\underline{q}:=(q_{i})_{i=-\infty}^{\infty},\epsilon)$ such that $p_i=q_i=0$ for $|i|>L,$ 
$q_i=p_i^{(\ell+1)}=p_i^{(\ell)}$ and
$q_i=q_i^{(\ell+1)}=q_i^{(\ell)}$ for $|i|\le L.$
For any $\epsilon\in [0,1]$, a fixed point $(\epsilon,\underline{p}=\underline{0}, \underline{q}=\underline{0})$ is called trivial. 
Trivial fixed points correspond to successful decoding. 
We define EBP EXIT-like curve  as the projected plots $(\epsilon, h^{\mathrm{EBP}}(\epsilon))$ of fixed points $(\epsilon,\underline{p},\underline{q})$, other than trivial ones, onto the following EXIT-like function. 
\begin{align*}
h^{\mathrm{EBP}}(\epsilon)=&f_{\epsilon,m}(z_i^{(\ell)})\cdot(z_i^{(\ell)})^{\dg}. 
\end{align*}

\begin{figure*}
\setlength{\unitlength}{1.0bp}%
\begin{picture}(400,175)(0,0)
\put(0,0)
{
\put(0,0){\includegraphics[width=0.5\textwidth]{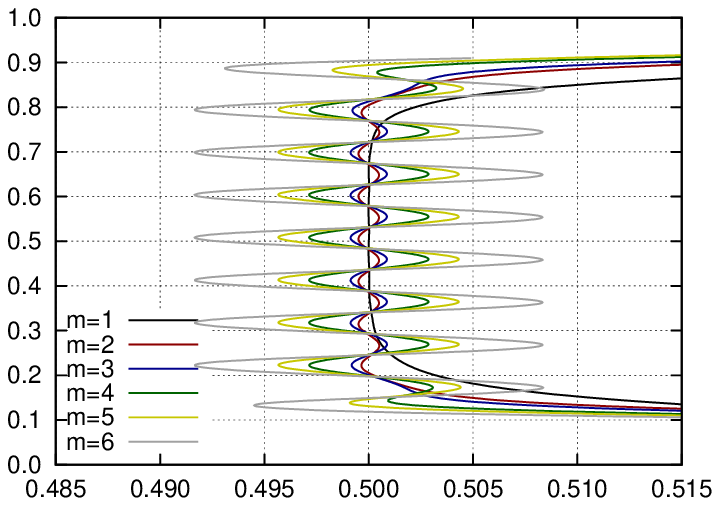} }
\put(260,0){\includegraphics[width=0.5\textwidth]{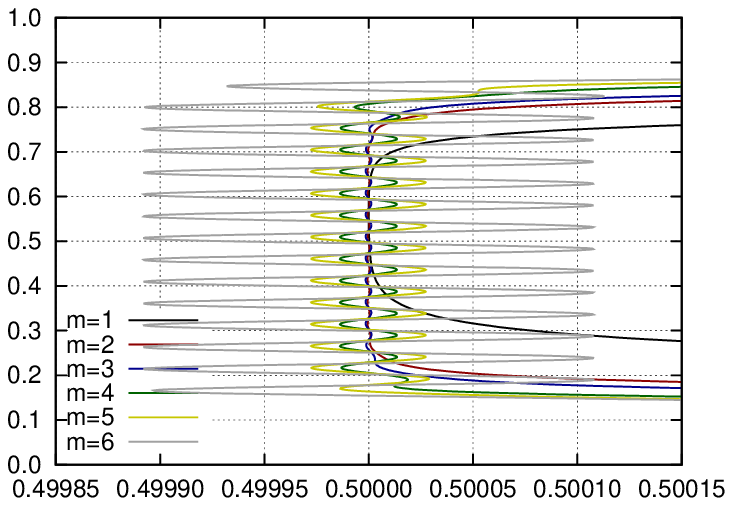} }
\put(0, 80){\rotatebox{90}{$h^{\mathrm{EBP}}(\epsilon)$}}
\put(260, 80){\rotatebox{90}{$h^{\mathrm{EBP}}(\epsilon)$}}
\put(130,0){$\epsilon$}
\put(370,0){$\epsilon$}
}
\end{picture}
\caption{Left: EBP EXIT-like  curve of SC-MN codes with $L=10, w=2$, right: EBP EXIT-like  curve of SC-MN codes with $L=20, w=3$, 
The channels are CD($m,\epsilon$)  for $m=1,\dotsc,6$.
The threshold for each code is given by the most left erasure probability $\epsilon$ on the EBP EXIT-like curve. 
Unfortunately and interestingly, wiggles are much amplified by increasing channel input size $2^m$. 
However, wiggles are significantly mitigated by increasing randomized window size $w$. Note that scale is magnified 100 times from left to right. }
\label{195320_3Feb12}
\end{figure*}

\begin{figure*}
\setlength{\unitlength}{1.0bp}%
\begin{picture}(400,170)(0,0)
\put(0,0)
{
\put(0,0){\includegraphics[width=0.5\textwidth]{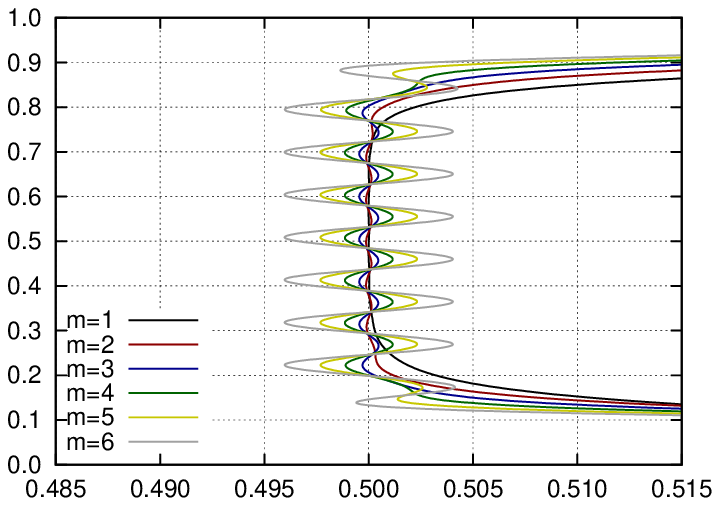} }
\put(260,0){\includegraphics[width=0.5\textwidth]{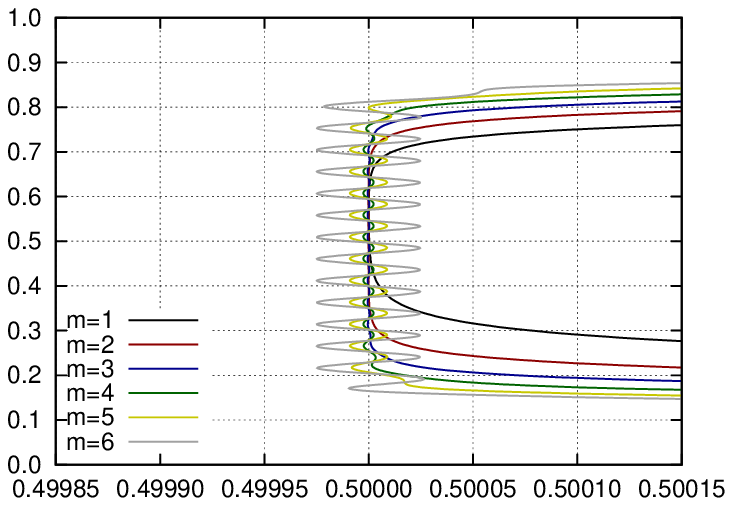} }
\put(0, 80){\rotatebox{90}{$h^{\mathrm{EBP}}(\epsilon)$}}
\put(260, 80){\rotatebox{90}{$h^{\mathrm{EBP}}(\epsilon)$}}
\put(130,0){$\epsilon$}
\put(370,0){$\epsilon$}
}
\end{picture}
\caption{Left: EBP EXIT-like curve of $(\dl=4, \dr=2, \dg=2, L, w)$ SC-MN codes with $L=10, w=2$, right: EBP EXIT-like curve of SC-MN codes with $L=20, w=3$, 
The channels are BD($m,\epsilon$) for $m=1,\dotsc,6$.
Wiggles are smaller than those of CD($m,\epsilon$) for the same $\dl, \dr, \dg, L$ and $w$. 
 }
\label{195418_3Feb12}
\end{figure*}

In Fig.~\ref{195027_3Feb12}, we plot the EBP EXIT-like curve of $(\dl=4, \dr=2, \dg=2, L, w=2)$ SC-MN codes for CD($m,\epsilon$). 
As can be seen, the threshold values are almost the same around $\epsilon=0.499509$ for $L=10$ and $L=20.$
In Figs.~\ref{195320_3Feb12} and \ref{195418_3Feb12}, we plot the EBP EXIT-like curve of $(\dl=4, \dr=2, \dg=2, L, w=2)$ SC-MN codes for CD($m,\epsilon$)  and BD($m,\epsilon$), respectively. From those curves, it can be seen that the threshold degradation is caused by large wiggles.  
The wiggle size seems to grow exponentially with $m$. 
This problem would be serious if $m$ got large. 
In fact, large $m$ is assumed in the network coding scenario \cite{DBLP:journals/corr/abs-0711-3935}. 
However, wiggles are significantly mitigated by increasing randomized window size $w$ as in the right column of Figs.~\ref{195320_3Feb12} and \ref{195418_3Feb12}.
%%%%%%%%%%%%%%%%%%%%%%%%%%%%%%%%%%%%%%%%%%%%%
% EBP CD
%%%%%%%%%%%%%%%%%%%%%%%%%%%%%%%%%%%%%%%%%%%%%
%%%%%%%%%%%%%%%%%%%%%%%%%%%%%%%%%%%%%%%%%%%%%
% EBP BEC
%%%%%%%%%%%%%%%%%%%%%%%%%%%%%%%%%%%%%%%%%%%%%
%%%%%%%%%%%%%%%%%%%%%%%%%%%%%%%%%%%%%%%%%%%%%%
\section{ Conclusion}
%%%%%%%%%%%%%%%%%%%%%%%%%%%%%%%%%%%%%%%%%%%%%%
In this paper, we evaluated the performance of binary SC-MN codes for the channels with affine space outputs. 
We derived DE equation and observed the threshold values are very close to the Shannon limit. 
We conclude that {\it binary} spatially-coupled codes are not only universal for binary input channels but likely universal also for non-binary input channels. 

The possible future works are (i) extension to AWGN  channels (ii) scaling the wiggle with $m$ and $w$ (iii) evaluating the performance for $m\to\infty$ (iv) 
decreasing the computational complexity $O(m^3)$ of channel detector (v) proving the capacity-achieving performance (vi)  extension to multiple access \cite{KuKaMAC} and memory \cite{KuKaDEC} channels. 
%%%%%%%%%%%%%%%%%%%%%%%%%%%%%%%%%%%%%%%%%%%%%%
% Bibliograpy
%%%%%%%%%%%%%%%%%%%%%%%%%%%%%%%%%%%%%%%%%%%%%%
\bibliographystyle{IEEEtran} 
\bibliography{IEEEabrv,../kenta_bib}
\end{document}